\documentclass[]{JHEP3}   % 10pt is ignored!
 \preprint{  }

\usepackage{epsfig,multicol}
\usepackage{amsmath}
\usepackage{graphics}
\usepackage{graphicx}
\usepackage{dcolumn}
\usepackage{amssymb}
\title{Entropy of the Schwarzschild-de Sitter Black Hole
due to arbitrary spin fields in different Coordinates}
\author{Chikun Ding and Jiliang  Jing\thanks{Corresponding author, Electronic address:
jljing@hunnu.edu.cn}\\
 Institute of Physics and  Department of Physics, \\
Hunan Normal University,\\ Changsha, Hunan 410081, P. R. China}

\abstract{

 By using the Newman-Penrose formalism and the improved thin-layer
``brick wall'' approach, the statistical-mechanical entropies of the
Schwarzschild-de Sitter black hole arising from quantum massless
arbitrary spin fields are studied in the Painlev\'e and Lemaitre
coordinates. Although the metrics in both the Painlev\'e and the
Lemaitre coordinates do not obviously possess the singularities as
that in the Schwarzschild-like coordinate, we find that, for
arbitrary spin fields, the entropies in the Painlev\'e and Lemaitre
coordinates are exactly equivalent to that in the Schwarzschild-like
coordinate. }


\begin{document}

\section{Introduction}

Bekenstein and Hawking \cite{be,ha} found that, by comparing black
hole physics with thermodynamics and from the discovery of black
hole evaporation, black hole entropy is proportional to the area
of the event horizon. The discovery is one of the most profound in
modern physics. However, the issue of the exact statistical origin
of the black hole entropy has remained a challenging one.
Recently, much effort has been concentrated on the problem
\cite{Carlip}-\cite{xli}. The ``brick wall'' model (BWM) proposed
by 't Hooft \cite{G.} is an extensively used way to calculate the
entropy in a variety of black holes, black branes, de Sitter
spaces, and anti-de Sitter spaces \cite{sqw}-\cite{xli}. In this
model the Bekenstein-Hawking entropy of the black hole is
identified with the statistical-mechanical entropy arising from a
thermal bath of quantum fields propagating outside the event
horizon. In refs. \cite{sqw, xli}, the authors found that although
the original BWM has contributed a great deal to the understanding
and calculation of the entropy of a black hole, there are some
drawbacks in it such as the little mass approximation and taking
the term including $L^3$ ($L$ being the ``infrared cutoff'') as a
contribution of the vacuum surrounding the black hole, etc. The
model is constructed on the basis of thermal equilibrium at a
large scale, so it cannot be applied to cases out of thermal
non-equilibrium problems, such as spacetime with two horizons, for
example, a Schwarzschild-de Sitter black hole and Vaidya black
hole \cite{cjg,xli}. Therefore, in this paper we utilize the
improved thin-layer BWM \cite{sqw,xli} to resolve Schwarzschild-de
Sitter spacetime.

In quantum field theory, we can use a timelike Killing vector to
define particle states. Therefore, in static spacetimes we know that
it is possible to define positive frequency modes by using the
timelike Killing vector. However, in these spacetimes there could
arise more than one timelike Killing vector which make the vacuum
states inequivalent. This means that the concept of particles is not
generally covariant and depends on the coordinate chosen to describe
the particular spacetime. It is therefore interesting to study the
following question: can we obtain the same statistical mechanical
entropies of the black hole in the Painlev\'e and the Lemaitre
coordinates due to the fact that in this question arises naturally
after Shankaranarayanan {\it et al.} \cite{sh1,sh2} studied the
Hawking temperature of the Schwarzschild black hole in the
Painlev\'e and the Lemaitre coordinates by using the method of
complex paths and they showed that the results are equal to those in
the Schwarzschild-like coordinate.  For the massless scalar field in
the general static black hole, Jing \cite{jing} find that the
entropies in the Painlev\'e and Lemaitre coordinates are exactly
equivalent to that in the Schwarzschild-like coordinate. However,
whether the entropies of the Schwarzschild-de Sitter black hole due
to arbitrary spin fields are the same in the Painlev\'e, the
Lemaitre and Schwarzschild-like coordinates is still an open
question. In this paper we will study the question carefully.

In order to compare the results obtained in this article with the
entropy of the Schwarzschild-de Sitter black hole in the
Schwarzschild-like coordinate, we first introduce the entropy for
the Schwarzschild-like coordinate. The Schwarzschild-de Sitter
spacetime in the Schwarzschild-like coordinate is described by
\begin{eqnarray}\label{metric1}
\label{3-1}ds^2=fdt_{s}^2-f^{-1}dr^2-r^2d\Omega^2,
\end{eqnarray}
 with
 \begin{eqnarray}
 f(r)=1-\frac{2m}{r}-\frac{\lambda}{3}r^2,\nonumber
 \end{eqnarray}
where $m$ is the mass of the black hole and $\lambda$ is
cosmological constant which we will assume it is positive, and the
geometric unit $G=c=\hbar=\kappa_B=1$ is used. The Schwarzschild-de
Sitter black hole have two horizons, i. e.,  the black hole event
horizon $r_H$ and the cosmological horizon $r_C$
\begin{eqnarray}
r_C=\frac{2}{\sqrt{\lambda}}\cos\alpha,
~~~r_H=\frac{2}{\sqrt{\lambda}}\cos(\alpha+\frac{\pi}{3}),~~~
\text{with}~~~\alpha=\frac{1}{3}\arctan\sqrt{\frac{1}{9\lambda
m^2}-1}.\nonumber
\end{eqnarray}

By using the improved thin-layer brick wall method, S. Q. Wu and M.
L. Yan \cite{sqw} found that the entropy of the Schwarzschild-de
Sitter black hole due to arbitrary spin field in the
Schwarzschild-like coordinate is
\begin{eqnarray}\label{entropy}
 S/g_s=\frac{15+(-1)^{2s}}{16}\Big[\frac{A_h}{48\pi\epsilon^2}
 +\frac{1}{45}(1-\frac{\lambda r_h^2}{2})\ln\frac{\Lambda}{\epsilon}\Big]
 -\frac{3+(-1)^{2s}}{4}\frac{\lambda(1+2s^2)}{36\pi}A_h\ln\frac{\Lambda}{\epsilon},
 \end{eqnarray}
where $g_s=1$ for scalar field ($s=0$), $g_s=2$ for Weyl neutrino
($s=1/2$), Maxwell electromagnetic ($s=1$), Rarita Schwinger
gravitino ($s=3/2$) and linearized Einstein gravitational ($s=2$)
fields, and $g_s=4$ for massless Dirac field ($s=1/2$), $A_h=4\pi
r_C^2$ or $4\pi r_H^2$, respectively. From (\ref{entropy}) we can
know that the entropies depend not only on the spins of the
particles but also on the cosmological constant except different
spin fields obey different statistics.

The paper is organized as follows. In Section 2 the metrics of the
Schwarzschild-de Sitter black hole in the Painlev\'e and Lemaitre
coordinates are introduced. In Section 3 the
 entropies of the Schwarzschild-de Sitter black
hole due to the arbitrary spin fields in the Painlev\'e and Lemaitre
coordinates are investigated. Section 4 is devoted to a summary.

\section{Metrics of Schwarzschild-de Sitter spacetime in
Painlev\'e and Lemaitre coordinates}

We now introduce the metrics of the Schwarzschild-de Sitter black
hole in the Painlev\'e and Lemaitre coordinates.

\subsection{Painlev\'e coordinate representation for
Schwarzschild-de Sitter black hole}

The time coordinate transformation from the Schwarzschild-like
coordinate (\ref{metric1}) to the Painlev\'e coordinate
\cite{jing} is
\begin{eqnarray}
  \label{sds}t=t_{s}+\int\frac{\sqrt{1-f(r)}}{f(r)}dr.
  \end{eqnarray}
    The radial and angular coordinates remain unchanged.
  With this transformation, the line element (\ref{metric1}) becomes
  \begin{eqnarray}
  \label{ds2}ds^2=g_{tt}dt^2+2g_{tr}dtdr+g_{rr}dr^2
  +g_{\theta\theta}d\theta^2+g_{\varphi\varphi}
  d\varphi^2,
\end{eqnarray}
with
 \begin{eqnarray}
 &&\label{g1}g_{tt}=f(r),~~
  g_{tr}=g_{rt}=-\sqrt{1-f(r)},~~g_{rr}=-1,\nonumber\\
  &&g_{\theta\theta}=-r^2,~~g_{\varphi\varphi}=-r^2\sin^2\theta.
  \end{eqnarray}
  The inverse metric is
  \begin{eqnarray}\label{g2}&&g^{tt}=1,~~g^{tr}=g^{rt}=
  -\sqrt{1-f(r)},\nonumber\\
  &&g^{rr}=-f(r),
  ~~~g^{\theta\theta}=-\frac{1}{r^2},~~g^{\varphi\varphi}=-\frac{1}{r^2\sin^2\theta
  }.
  \end{eqnarray}

  The metric in the Painlev\'e coordinate has distinguishing features:
(a) The spacetime is stationary but not static, so the time
direction remains to be a Killing vector; (b) there is now no
singularity at $f(r)=0$, so the metric is regular at horizons of the
black hole.  That is to say, the coordinate complies with
perspective of a free-falling observer, who is expected  to
experience nothing out of the ordinary upon passing through the
event horizon. However, we will see next that the event and
cosmological horizons manifests themselves as  singularities in the
expression for the semiclassical action.

\subsection{Lemaitre coordinate representation for Schwarzschild-de Sitter black hole}

~~~~The coordinates that transform from the Painlev\'e
   coordinate (\ref{ds2}) to the Lemaitre coordinate
   $(V,~U,~\theta,~\varphi)$ are given by
   \begin{eqnarray}\label{lem}
U=\tilde{r}-t,~~~V=\tilde{r}+t,
   \end{eqnarray}
   with
\begin{eqnarray}
   \tilde{r}=t+\int\frac{dr}{\sqrt{1-f(r)}},\nonumber
   \end{eqnarray}
   where $t$ is the Painlev\'e time and $V$ is the Lemaitre time. The angular
   coordinates $\theta$ and $\varphi$ remain the same. The line element
   (\ref{ds2}) in the new coordinate becomes
\begin{eqnarray}\label{Lm}
ds^2=g_{VV}dV^2+2g_{VU}dVdU+g_{UU}dU^2
  +\tilde{g}_{\theta\theta}d\theta^2+\tilde{g}_{\varphi\varphi}
  d\varphi^2,
\end{eqnarray}
with
 \begin{eqnarray}
 &&\label{g3}g_{VV}=g_{UU}=\frac{1-\tilde{f}}{4},
  ~~g_{VU}=g_{UV}=-\frac{\tilde{f}+1}{4},\nonumber\\
  &&\tilde{g}_{\theta\theta}=-y,~~\tilde
  {g}_{\varphi\varphi}=-y\sin^2\theta,
  \end{eqnarray}
  where
  \begin{eqnarray}
\tilde{f}(U)=1-f(r),~~y(U)= r^2.\nonumber
\end{eqnarray}
  The inverse metric is
  \begin{eqnarray}\label{g4}&&g^{VV}=g^{UU}=-\frac{1-\tilde{f}}{\tilde{f}},~~g^{VU}=g^{UV}
  =-\frac{\tilde{f}+1}{\tilde{f}},\nonumber\\
  &&
  \tilde{g}^{\theta\theta}=-\frac{1}{y},~~\tilde{g}^{\varphi\varphi}=-\frac{1}{y\sin^2\theta
  }.
  \end{eqnarray}
We can see that the Lemaitre coordinate is time-dependant and the
metric (\ref{Lm}) has no coordinate singularity just as in the
Painlev\'e coordinates. However, we will know that the event and
cosmological horizons also manifests themselves as  singularities in
the expression for the semiclassical action.
\section{Entropy of Schwarzschild-de Sitter black hole due to arbitrary spin fields in different coordinates}

In this section we will study the entropy of the Schwarzschild-de
Sitter black hole due to arbitrary spin fields in the Painlev\'e
and Lemaitre coordinates by using the improved thin-layer brick
wall method.

\subsection{Entropy of Schwarzschild-de Sitter black hole due to arbitrary spin fields in Painlev\'e
coordinate}

Now in order to derive the master equation for arbitrary spin
fields in Painlev\'e coordinate (\ref{ds2}), we work it within the
Newman-Penrose formalism \cite{en,sch} by taking covariant
components of the null tetrad vectors as

\begin{eqnarray}
&&\label{te1}l_\mu=\Big(1,~-\frac{1+\sqrt{1-f(r)}}{f(r)},~0,~0\Big),
~~~~n_\mu=\frac{1}{2}\Big(f(r),~\frac{f(r)}{1+\sqrt{1-f(r)}},~0,~0\Big),\nonumber\\
&&m_\mu=-\frac{r}{\sqrt{2}}\big(0,~0,~1,~i\sin\theta\big),~~~~~~~~~~~~\overline
m_\mu=-\frac{r}{\sqrt{2}}\big(0,~0,~1,~-i\sin\theta\big),
\end{eqnarray}
The non-zero spin coefficients are
\begin{eqnarray}\label{rot}
&&\rho=-\frac{1}{r},\nonumber\\
&&\mu=-\frac{1}{2r}f(r),\nonumber\\
&&\gamma=\frac{1
}{4}f'(r),\nonumber\\
&&\alpha=-\beta=-\frac{1}{2\sqrt{2}r}\cot\theta,
\end{eqnarray}
where a prime denotes the differential with respect to $r$, and only
one non-zero Weyl tensor is

\begin{eqnarray}\label{psi}
\Psi_2=-\frac{m}{r^3}+\frac{\lambda}{3}.
\end{eqnarray}

 Assuming that the azimuthal and time dependence of the
perturbed fields will be of the form $e^{i(m\varphi-E t)}$, we
find that the directional derivatives are
\begin{eqnarray}\label{div}
&&D=l^\mu\partial_\mu=\mathcal{D}_0,~~~~~~~~~~\Delta=n^\mu\partial_\mu=-\frac{\Delta_r}{2r^2}\mathcal{D}^\dagger_0,\nonumber\\
&&\delta=m^\mu\partial_\mu=\frac{1}{\sqrt{2}r}\mathcal{L}^\dag_0,~~~~\bar\delta=\bar
m^\mu\partial_\mu=\frac{1}{\sqrt{2}r}\mathcal{L}_0,
\end{eqnarray}
with
\begin{eqnarray}\label{dl}
&&\mathcal{D}_n=\frac{\partial}{\partial
r}-\frac{iK_1}{\Delta_r}\big(1+\sqrt{1-f(r)}\big)+n\frac{\Delta'_r}{\Delta_r},
\nonumber\\
&&\mathcal{L}_n=\frac{\partial}{\partial\theta}-K_2+n\cot\theta,\nonumber\\
&&\mathcal{D}_n^\dag=\frac{\partial}{\partial
r}+\frac{iK_1}{\Delta_r}\big(1-\sqrt{1-f(r)}\big)+n\frac{\triangle_r'}{\Delta_r},
\nonumber\\
&&\mathcal{L}_n^\dag=\frac{\partial}{\partial\theta}+K_2+n\cot\theta,
\end{eqnarray}
where
\begin{eqnarray}
\Delta_r=r^2 f(r),~~K_1=E r^2,~~K_2=-\frac{m}{\sin\theta}.
\end{eqnarray}

With the help of the Newman-Penrose formalism \cite{en,sch},  it can
be shown that decoupled master equations controlling the
perturbations of Schwarzschild-de Sitter black hole for massless
arbitrary spin fields (i.e., scalar, Weyl neutrino, source-free
Maxwell electromagnetic, Rarita-Schwinger gravitino, and the
linearized Einstein gravitational fields) read \cite{hsu} \cite{sat}
\cite{gft}
\begin{eqnarray}\label{fie1}
&&\{[D-(2s-1)\epsilon+\epsilon^*-2s\rho-\rho^*](\Delta-2s\gamma+\mu)-[\delta-(2s-1)\beta-\alpha^*](\bar
\delta-2s\alpha)\nonumber\\
&&-(s-1)(2s-1)\Psi_2\}\Phi_s=0,
\end{eqnarray}
for $s=1/2,1,3/2,2$ and
\begin{eqnarray}\label{fie2}
&&\{[\Delta+(2s-1)\gamma-\gamma^*+2s\mu+\mu^*](D+2s\epsilon-\rho)-[\bar\delta+(2s-1)\alpha+\beta^*](
\delta+2s\beta)\nonumber\\
&&-(s-1)(2s-1)\Psi_2\}\Phi_{-s}=0,
\end{eqnarray}
for $s=0,~-1/2,~-1,~-3/2,~-2$.

Using Eqs. (\ref{rot}), (\ref{psi}) and (\ref{div}), Eqs.
(\ref{fie1}) and (\ref{fie2}) can be expressed as
\begin{eqnarray}\label{partial}
&&\frac{1}{r^2}\Big\{\Delta_r^{-s}\frac{\partial}{\partial
r}\big(\Delta_r^{1+s}\frac{\partial}{\partial
r}\big)-2iK_1\sqrt{1-f(r)}\frac{\partial}{\partial
r}+\frac{f(r)K_1^2-isK_1\Delta_r'(1+\sqrt{1-f(r)})}{\Delta_r}
\nonumber\\&&+\frac{s}{2}\Delta_r''+\frac{1}
{\sin\theta}\frac{\partial}{\partial\theta}
\big(\sin\theta\frac{\partial}{\partial\theta}\big)-[K_2-s\cot\theta]^2
+4is\omega r-(4s^2+2)\frac{\lambda
r^2}{3}\Big\}\Phi_s=0,\nonumber\\&&
\frac{1}{r^2}\Big\{\Delta_r^{s}\frac{\partial}{\partial
r}\big(\Delta_r^{1-s}\frac{\partial}{\partial
r}\big)-2iK_1\sqrt{1-f(r)}\frac{\partial}{\partial
r}+\frac{f(r)K_1^2+isK_1\Delta_r'(1+\sqrt{1-f(r)})}{\Delta_r}
\nonumber\\&&-\frac{s}{2}\Delta_r''+\frac{1}{\sin\theta}\frac{\partial}{\partial\theta}
\big(\sin\theta\frac{\partial}{\partial\theta}\big)-[K_2+s\cot\theta]^2
-4is\omega r-(4s^2+2)\frac{\lambda r^2}{3}\Big\}(r^{2s}\Phi_{-s})=0.
\end{eqnarray}
We can easily find that they are dual by interchanging $s=-s$. Thus
one only needs to consider the case of positive spin state $s$, and
obtain the results for the negative spin state $-s$ by substituting
$s\longrightarrow-s$. Two of equations (\ref{partial}) can be
combined into the form of Teukolsky's master equation \cite{sat}
\begin{eqnarray}\label{teu}
&&\Big\{f(r)\frac{\partial^2}{\partial
r^2}+\frac{(1+s)\triangle_r'}{r^2}\frac{\partial}{\partial
r}-2iE\sqrt{1-f(r)}\frac{\partial}{\partial
r}\nonumber\\
&&+\frac{1}{r^2}\frac{\partial^2}{\partial\theta^2}
+\frac{\cot\theta}{r^2}\frac{\partial}{\partial\theta}+E^2
-\frac{m^2}{r^2\sin^2\theta}-\frac{isE\Delta_r'}{\Delta_r}
(1+\sqrt{1-f(r)})\nonumber\\
&&+\frac{4isE}{r}-\frac{2sm\cot\theta}{r^2\sin\theta}
+\frac{s}{2r^2}\Delta_r''-\frac{\lambda}{3}(4s^2+2)
-\frac{s^2}{r^2}\cot^2\theta\Big\}\tilde{\Phi}_s=0.
\end{eqnarray}

Now we can calculate the entropy due to arbitrary spin fields for
the nonextreme Schwarzschild-de Sitter black hole in Painlev\'e
coordinate by the thin-layer BWM. First we try to seek the total
number of modes with energy less than $E$. In order to do this, we
make use of the WKB approximation and substitute $\tilde{\Phi}_s\sim
e^{iG(r,~\theta)}$ into the above Teukolsky's master Eq.
(\ref{teu}), then we obtain
\begin{eqnarray}\label{krk}
&&f(r)k_r^2-2E\sqrt{1-f(r)}k_r+\frac{1}{r^2}k_\theta^2-E^2+\frac{m^2}{r^2\sin^2\theta}
+\frac{2sm}{r^2\sin\theta}\cot\theta\nonumber\\
&&+(\frac{s\cot\theta}{r})^2
+\frac{\lambda}{3}(4s^2+2)-\frac{s}{2r^2}\Delta_r''=0,
\end{eqnarray}
where $k_r=G,_r$ and $k_\theta=G,_\theta$ are the momentum of the
particles moving in $r$ and $\theta$, respectively. In terms of the
covariant metric components $g_{\mu\nu}$, Eq. (\ref{krk}) can be
rewritten as
\begin{eqnarray}\label{krk2}
f(r)k_r^2-2E\sqrt{1-f(r)}k_r-E^2-\frac{k_\theta^2}
{g_{\theta\theta}}-\frac{m^2}{g_{\varphi\varphi}}+H_s=0,
\end{eqnarray}
where
\begin{eqnarray}
 H_s=\frac{2sm}{r^2\sin\theta}\cot\theta+\frac{s^2}
 {r^2}\cot^2\theta
+\frac{\lambda}{3}(4s^2+2)-\frac{s}{2r^2}\Delta_r''.
\end{eqnarray}
The roots of the Eq. (\ref{krk2}) are
\begin{eqnarray}
k_r^\pm=\frac{E\sqrt{1-f(r)}\pm\sqrt{E^2-f(r)
\big[-\frac{k_\theta^2}{g_{\theta\theta}}-\frac{m^2}
{g_{\varphi\varphi}}+H_s\big]}}{f(r)},
\end{eqnarray}
the sign ambiguity of the square root is related to the ``out-going"
($k_r^+$) or ``in-going" ($k_r^-$) particles, respectively. Here we
utilize the average of the radial momentum (the minus before the
$k_r^-$ is caused by a different direction),
\begin{eqnarray}\label{kraverage}
 \tilde{k}_r=\frac{k_r^+-k_r^-}{2}
 =\frac{1}{f(r)}\sqrt{E^2-f(r)\Big(-\frac{k_\theta^2}
 {g_{\theta\theta}}-\frac{m^2}{g_{\varphi\varphi}}+H_s\Big)}.
\end{eqnarray}
So in this way, we take all kinds of particles into account. Eq.
(\ref{kraverage}) can be rewritten as
\begin{eqnarray}
 \tilde{k}_r=\frac{1}{\sqrt{-g^{rr}}}
 \sqrt{\frac{E^2}{g_{tt}}-\Big[-\frac{k_\theta^2}{g_{\theta\theta}}
 -\frac{(m+m_0)^2}{g_{\varphi\varphi}}+V_s\Big]},
\end{eqnarray}
with
\begin{eqnarray}
m_0= s\cos\theta,
~~V_s=\frac{\lambda}{3}(4s^2+2)-\frac{s}{2r^2}\Delta_r''.
\end{eqnarray}
 The number of modes with $E$ is equal to the number of
states in this classical phase space \cite{mhl}
\begin{eqnarray}\label{n1}
n_h(E,s)&=&\frac{1}{(2\pi)^3}\int drd\theta d\varphi\int
d\tilde{k}_rdk_\theta dm\nonumber\\&=&\frac{1}{3\pi}\int
d\theta\int^{r_h+N\varepsilon}_{r_h+\varepsilon}dr
\frac{\sqrt{-g}}{(g_{tt})^2}[E^2-g_{tt}V_s]^{3/2},
\end{eqnarray}
under the improved thin-layer BWM boundary conditions
\begin{eqnarray} &&\Phi(t,r,\theta,\varphi)=0 ~~~\text{for}~~~
r< r_H+\varepsilon~~~~~\text{and}~~~ r> r_H+N\varepsilon,\nonumber \\
&&\Phi(t,r,\theta,\varphi)=0 ~~~\text{for}~~~ r<
r_C-N\varepsilon~~~\text{and}~~~r> r_C-\varepsilon,\nonumber
\end{eqnarray}
where $\varepsilon\ll r_H$ (or $r_C$), $N$ is an arbitrary big
integer which removes the infrared divergence. It is obviously that
the location of the brick wall and the meaning of this wall in the
Painlev\'e coordinate are the same as that in the Schwarzschild-like
coordinate.

The integral is taken only over those values for which the square
root in Eq. (\ref{n1}) exists. Summing over the positive and
negative spin states $\pm s$, we get the total states number
\begin{eqnarray}\label{gama}
n_h(E)=\frac{g_s}{2}[n_h(E, s)+n_h(E,
-s)]\approx\frac{g_s}{3\pi}[I_{1h}E^3+3I_{2h}E],
\end{eqnarray}
where $I_{1h}$ ($I_{2h}$) represents $I_{1H}$ ($I_{2H}$) for event
horizon or $I_{1C}$ ($I_{2C}$) for the cosmological horizon, and
these quantities are given by
\begin{eqnarray}&&I_{1H}=\int
d\theta\int^{r_H+N\varepsilon}_{r_H+\varepsilon}dr
\frac{\sqrt{-g}}{g^2_{tt}},~~~~~~~~~~~~~~~~~~~~~ I_{1C}=\int
d\theta\int^{r_C-\varepsilon}_{r_C-N\varepsilon}dr
\frac{\sqrt{-g}}{g^2_{tt}},\nonumber\\
&&I_{2H}=\int d\theta\int^{r_H+N\varepsilon}_{r_H+\varepsilon}dr
\frac{\sqrt{-g}}{g_{tt}}\big[\frac{\lambda}{3}(2s^2+1)\big],~~
~~I_{2C}=\int d\theta\int^{r_C-\varepsilon}_{r_C-N\varepsilon}dr
\frac{\sqrt{-g}}{g_{tt}}\big[\frac{\lambda}{3}(2s^2+1)\big].\nonumber
\end{eqnarray}
 In the above, we have expanded Eq. (\ref{gama}) in the
high frequency approximation and introduced an appropriate
degeneracy $g_s$ for each species of particles. Accordingly, the
free energy $F$ at inverse Hawking temperature $\beta$ can be
expressed as

\begin{eqnarray}
F_h&=&-\int_0^\infty
dE\frac{n_h(E)}{e^{\beta E}-(-1)^{2s}}\nonumber\\
&=&-g_s\Big[2\zeta(4)\frac{15+(-1)^{2s}}{16\pi\beta^4}I_{1h}
+\zeta(2)\frac{3+(-1)^{2s}}{4\pi\beta^2}I_{2h}\Big],
\end{eqnarray}
where $\zeta(n)=\Sigma^\infty_{k=1}1/k^n$ is the Riemann zeta
function, $\zeta(4)=\pi^4/90,~\zeta(2)=\pi^2/6$, etc. We can now
 obtain the entropy of the Schwarzschild-de Sitter black
hole due to arbitrary spin fields from the standard formula
$S_h=\beta^2(\partial F_h/\partial\beta)$
\begin{eqnarray}\label{entropy2}
S_h/g_s=\frac{15+(-1)^{2s}}{16}\Big[\frac{A_h}{48\pi\epsilon_h^2}
 +\frac{1}{45}(1-\frac{\lambda r_h^2}{2})\ln\frac{\Lambda_h}
 {\epsilon_h}\Big]
 -\frac{3+(-1)^{2s}}{4}\frac{\lambda(1+2s^2)}{36\pi}A_h
 \ln\frac{\Lambda_h}{\epsilon_h},
\end{eqnarray}
where the ultraviolet cutoff $\epsilon_h$ and the infrared cutoff
$\Lambda_h$ have been set by $\eta_h^2=2\epsilon_h^2/15$ and
$N=\Lambda_h^2/\epsilon_h^2$ \cite{rbm,jlj}, the proper distance
$\eta_h$ from the event horizon to the inner brick wall is
$\eta_H=\int_{r_H}^{r_H+\varepsilon}\sqrt{-g_{rr}
+g^2_{tr}/g_{tt}}dr\thickapprox 2r_H(\varepsilon
/\Delta'_{r_H})^{1/2}=2\sqrt{\varepsilon r_H/(1-\lambda r_H^2)}$ and
from the cosmological horizon to the brick wall is
$\eta_C=\int^{r_C}_{r_C-\varepsilon}\sqrt{-g_{rr}+g^2_{tr}
/g_{tt}}dr\thickapprox 2\sqrt{\varepsilon r_C/(1-\lambda r_C^2)}$,
and $A_h=4\pi r_H^2$ or $4\pi r_C^2$.

 We find
that Eq. (\ref{entropy2}) is in agreement with Wu-Yan's result
(\ref{entropy}).
 That is to say, the entropy calculated in the Painlev\'e
 coordinate is exactly equal to that in the
 Schwarzschild-like coordinate.

By the equivalence principle and the standard
 quantum field theory in flat space,
  to construct  a vacuum state for the massless scalar field in the
  Painlev\'e spacetime we should leave all the positive
  frequency modes empty. Kraus \cite{pkra} pointed out that for the
  metric (\ref{ds2}) it is convenient to work along a curve
  $dr+\sqrt{1-g(r)}dt=0$, then the condition is simply a positive
  frequency with respect to $t$ near this curve. It is easy to
  prove that the modes used to calculate the entropy are
  essentially the same as that in the Schwarzschild-like
  coordinate.

\subsection{ Entropy of Schwarzschild-de Sitter black hole due to arbitrary spin fields in Lemaitre
coordinate}

 Now we calculated the entropy of Schwarzschild-de Sitter black
hole due to arbitrary spin field in Lemaitre coordinates.

For the metric (\ref{Lm}), The null tetrad vectors can be
expressed as
\begin{eqnarray}
&&\label{te3}l_\mu=\Big(-\frac{1}{2}\sqrt{\tilde{f}},~\frac{\sqrt{\tilde{f}}(1-\sqrt{\tilde{f}})}{2(1+\sqrt{\tilde{f}})},~0,~0\Big),
~~n_\mu=\Big(-\frac{(1-\tilde{f})}{4\sqrt{\tilde{f}}},~\frac{(1+\sqrt{\tilde{f}})^2}{4\sqrt{\tilde{f}}},~0,~0\Big),\nonumber\\
&&m_\mu=-\frac{\sqrt{y}}{\sqrt{2}}\big(0,~0,~1,~i\sin\theta\big),~~~~~~~~~~~~~\overline
m_\mu=-\frac{\sqrt{y}}{\sqrt{2}}\big(0,0,1,-i\sin\theta\big).
\end{eqnarray}
 We find the non-zero spin
coefficients
\begin{eqnarray}\label{spin2}
&&\rho=-\frac{1}{\sqrt{y}},\nonumber\\
&&\mu=\frac{(1-\tilde{f})}{\sqrt{\tilde{f}}}\frac{2}{\sqrt{y}},
\nonumber\\
&&\gamma=\frac{2}{\sqrt{y}}-\frac{2\lambda\sqrt{y}}{\tilde{f}}+2(\frac{m}{y}-\frac{\lambda}{3}\sqrt{y}),\nonumber\\
&&
\epsilon=-\frac{2}{\sqrt{\tilde{f}}}(\frac{m}{y}-\frac{\lambda}{3}\sqrt{y}),\nonumber\\
&&\alpha=-\beta=-\frac{1}{2\sqrt{2y}}\cot\theta,
\end{eqnarray}
and only one non-zero Weyl tensor
\begin{eqnarray}
\nonumber \Psi_2=-\frac{m}{y\sqrt{y}}+\frac{\lambda}{3}.
\end{eqnarray}

Assuming that the azimuthal and time dependence of the perturbed
fields will be of the form $e^{i(m\varphi-E V)}$, we find that the
directional derivatives are
\begin{eqnarray}\label{derive2}
&&D=l^\mu\partial_\mu=\mathcal{D}_0,~~~~~~~~~~\Delta=n^\mu\partial_\mu=-\frac{\Delta_U}{2y}\mathcal{D}^\dagger_0,\nonumber\\
&&\delta=m^\mu\partial_\mu=\frac{1}{\sqrt{2y}}\mathcal{L}^\dag_0,~~
~~\bar\delta=\bar
m^\mu\partial_\mu=\frac{1}{\sqrt{2y}}\mathcal{L}_0,
\end{eqnarray}
with
\begin{eqnarray}
&&\mathcal{D}_n=\frac{\partial}{\partial
U}-\frac{iK_1}{\Delta_U}(1-\sqrt{\tilde{f}})^2+n\frac{\Delta'_U}{\Delta_U},
\nonumber\\
&&\mathcal{L}_n=\frac{\partial}{\partial\theta}-K_2+n\cot\theta,\nonumber\\
&&\mathcal{D}_n^\dag=\frac{1}{\tilde{f}}\frac{\partial}{\partial
U}-\frac{iK_1}{\triangle_U}\frac{(1+\sqrt{\tilde{f}})^2}{\tilde{f}}+n\frac{\triangle_U'}{\Delta_U},
\nonumber\\
&&\mathcal{L}_n^\dag=\frac{\partial}{\partial\theta}+K_2+n\cot\theta,
\end{eqnarray}
where
\begin{eqnarray}
\Delta_U=y(1-\tilde{f}),~~K_1=E y,~~K_2=-\frac{m}{\sin\theta}.
\end{eqnarray}
Substituting (\ref{derive2}) and (\ref{spin2}) into (\ref{fie1})
and (\ref{fie2}), we can obtain the Teukolsky's master equation
\begin{eqnarray}\label{teu2}
&&\Big\{(1-\tilde{f})\frac{1}{\tilde{f}}\frac{\partial^2}{\partial
U^2}+\frac{(1+s)\triangle_U'}{y}\frac{\partial}{\partial
U}-\frac{(1+\tilde{f})2iE}{\tilde{f}}\frac{\partial}{\partial
U}+\frac{1}{y}\frac{\partial^2}{\partial\theta^2}
+\frac{\cot\theta}{y}\frac{\partial}{\partial\theta}-\frac{1-\tilde{f}}{\tilde{f}}E^2
-\frac{m^2}{y\sin^2\theta}\nonumber\\
&&-\frac{isE\Delta_U'}{\Delta_U}(1-\sqrt{\tilde{f}})^2+\frac{4isE}{\sqrt{y}}-\frac{2sm\cot\theta}{y\sin\theta}
+\frac{s}{2y}\Delta_U''-\frac{\lambda}{3}(4s^2+2)-\frac{s^2}{y}\cot^2\theta\Big\}\tilde{\Phi}_s=0.
\end{eqnarray}

 Taking $\tilde{\Phi}_s\sim
e^{iG(U,~\theta)}$ into the above Teukolsky's master equation
(\ref{teu2}), we have
\begin{eqnarray}
\frac{1-\tilde{f}}{\tilde{f}}k_U^2-\frac{1+\tilde{f}}{\tilde{f}}2E
k_U+\frac{1-\tilde{f}}{\tilde{f}}E^2-\frac{k_\theta^2}{\tilde{g}_{\theta\theta}}-\frac{m^2}{\tilde{g}_{\varphi\varphi}}+H_s=0,
\end{eqnarray}
with
\begin{eqnarray}
 H_s=\frac{2sm}{y\sin\theta}\cot\theta+\frac{s^2}{y}\cot^2\theta
+\frac{\lambda}{3}(4s^2+2)-\frac{s}{2y}\Delta_U'',
\end{eqnarray}
where $k_U=G,_U$ and $k_\theta=G,_\theta$ are the momentum of the
particle moving in $U$ and $\theta$, respectively. We get the
roots of $k_U$ as
\begin{eqnarray}
k_U^\pm=\frac{1+\tilde{f}}{(1-\tilde{f})}E\pm\sqrt{\frac{\tilde{f}}{1-\tilde{f}}}\sqrt{\frac{4E^2}{1-\tilde{f}}
+\frac{k_\theta^2}{\tilde{g}_{\theta\theta}}+\frac{m^2}{\tilde{g}_{\varphi\varphi}}-H_s},
\end{eqnarray}
the roots are related to the ``out-going" ($k_U^+$) and ``in-going"
($k_U^-$) particles, respectively.

Here, we make use of the average of the $U$-direction momentum
(the minus before the $k_U^-$ is caused by a different direction)
\begin{eqnarray}\label{ku}
\tilde{k}_U=\frac{k_U^+-k_U^-}{2}=\sqrt{\frac{\tilde{f}}{1-\tilde{f}}}\sqrt{\frac{4E^2}{1-\tilde{f}}
+\frac{k_\theta^2}{\tilde{g}_{\theta\theta}}+\frac{m^2}{\tilde{g}_{\varphi\varphi}}-H_s}.
\end{eqnarray}
Eq. (\ref{ku}) can be rewritten as
\begin{eqnarray}
\tilde{k}_U=\frac{1}{\sqrt{-g^{UU}}}\sqrt{\frac{E^2}{g_{VV}}
+\frac{k_\theta^2}{\tilde{g}_{\theta\theta}}+\frac{(m+m_0)^2}
{\tilde{g}_{\varphi\varphi}}-V_s},
\end{eqnarray}
with\begin{eqnarray}
m_0=s\cos\theta,~~V_s=\frac{\lambda}{3}(4s^2+2)-\frac{s}{2y}\Delta''_U.
\end{eqnarray}
 Summing over the positive and negative spin states
$\pm s$, we get the total states number
\begin{eqnarray}\label{gama2}
n_h(E)=\frac{g_s}{2}[n_h(E, s)+n_h(E,
-s)]\approx\frac{g_s}{3\pi}[I_{1h}E^3+3I_{2h}E],
\end{eqnarray}
with
\begin{eqnarray}\nonumber
&&I_{1H}=\int
d\theta\int^{U_H+\tilde{N}\tilde{\varepsilon}}_{U_H+\tilde{\varepsilon}}dU
\frac{\sqrt{-\tilde{g}}}{(g_{VV})^2},~~~~~~~~~~~~~~~~I_{1C}=\int
d\theta\int_{U_C-\tilde{N}\tilde{\varepsilon}}^{U_C-\tilde{\varepsilon}}dU
\frac{\sqrt{-\tilde{g}}}{(g_{VV})^2},\nonumber\\
&&I_{2H}=\int
d\theta\int^{U_H+\tilde{N}\tilde{\varepsilon}}_{U_H+\tilde{\varepsilon}}dU
\frac{\sqrt{-\tilde{g}}}{g_{VV}}\big[\frac{\lambda}{3}(2s^2+1)\big],~~
I_{2C}=\int
d\theta\int_{U_C-\tilde{N}\tilde{\varepsilon}}^{U_C-\tilde{\varepsilon}}dU
\frac{\sqrt{-\tilde{g}}}{g_{VV}}\big[\frac{\lambda}{3}(2s^2+1)\big],
\end{eqnarray}
where\begin{eqnarray}&&\sqrt{-\tilde{g}}=\frac{1}{2}\sin\theta
\sqrt{\frac{3m^2}{\lambda}\big[\frac{1-2e^{\sqrt{3\lambda}U}+e^{2\sqrt{3\lambda}U}}{e^{\sqrt{3\lambda}U}}\big]
+\frac{\lambda}{3}\big[\frac{3m(1-2e^{\sqrt{3\lambda}U}+e^{2\sqrt{3\lambda}U})}
{2\lambda e^{\sqrt{3\lambda}U}}\big]^2},\nonumber\\
&&g_{VV}=\frac{1}{4}\Big\{1-2m[\frac{3m(1-2e^{\sqrt{3\lambda}U}+e^{2\sqrt{3\lambda}U})}{2\lambda
e^{\sqrt{3\lambda}U}}\big]^{-1/3}
-\frac{\lambda}{3}[\frac{3m(1-2e^{\sqrt{3\lambda}U}+e^{2\sqrt{3\lambda}U})}
{2\lambda e^{\sqrt{3\lambda}U}}\big]^{2/3}\Big\}.\nonumber
\end{eqnarray}
In above calculation, we used the improved thin-layer BWM boundary
conditions
\begin{eqnarray}
&&\Phi(V,U,\theta,\varphi)=0 ~~~\text{for}~~~ U<
U_H+\tilde{\varepsilon}~~~~~\text{and}~~~ U>
U_H+\tilde{N}\tilde{\varepsilon},\nonumber
\\
&&\Phi(V,U,\theta,\varphi)=0 ~~~\text{for}~~~U<
U_C-\tilde{N}\tilde{\varepsilon}~~~\text{and}~~~U>
U_C-\tilde{\varepsilon},\nonumber
\end{eqnarray}where
$\tilde{\varepsilon}=2/\sqrt{3\lambda}\ln\big[\sqrt{\lambda
\varepsilon^3/6m}+\sqrt{\lambda\varepsilon^3/6m+1}\big]$ which gives
the relation between the location of the brick wall in the Lemaitre
and Schwarzschild-like coordinates, $\tilde{N}$ is an arbitrary big
integer, and $U_H$ and $U_C$ are
\begin{eqnarray}
U_H&=&\frac{2}{\sqrt{3\lambda}}\ln\Big[
\sqrt{\frac{4\cos^3(\alpha+\pi/3)}{\cos3\alpha}}
+\sqrt{\frac{4\cos^3(\alpha+\pi/3)}{\cos3\alpha}+1}\Big]\nonumber\\
   &=&\frac{2}{\sqrt{3\lambda}}\ln\Big[\sqrt{\frac{\lambda
r_H^3}{6m}}+\sqrt{\frac{\lambda r_H^3}{6m}+1}\big],\nonumber\\
U_C&=&\frac{2}{\sqrt{3\lambda}}\ln\Big[\sqrt{\frac{4\cos^3\alpha}{\cos3\alpha}}
+\sqrt{\frac{4\cos^3\alpha}{\cos3\alpha}+1}\Big]\nonumber\\
&=&\frac{2}{\sqrt{3\lambda}}\ln\Big[\sqrt{\frac{\lambda
r_C^3}{6m}}+\sqrt{\frac{\lambda r_C^3}{6m}+1}\big],\nonumber
\end{eqnarray}
which correspond to the event and cosmological horizons of the
Schwarzschild-de Sitter black hole.

 Then, the free energy can be expressed as
\begin{eqnarray}
F_h&=&-\int_0^\infty
dE\frac{n_h(E)}{e^{\beta E}-(-1)^{2s}}\nonumber\\
&=&-g_s\Big[2\zeta(4)\frac{15+(-1)^{2s}}{16\pi\beta^4}I_{1h}
+\zeta(2)\frac{3+(-1)^{2s}}{4\pi\beta^2}I_{2h}\Big],
\end{eqnarray}
 We can now
 obtain the entropy of the Schwarzschild-de Sitter black
hole due to arbitrary spin fields in Lemaitre coordinate as
\begin{eqnarray}\label{entropy3}
S_h/g_s=\frac{15+(-1)^{2s}}{16}\Big[\frac{A_h}{48\pi\epsilon_h^2}
 +\frac{1}{45}(1-\frac{\lambda r_h^2}{2})\ln\frac{\Lambda_h}
 {\epsilon_h}\Big]
 -\frac{3+(-1)^{2s}}{4}\frac{\lambda(1+2s^2)}{36\pi}A_h
 \ln\frac{\Lambda_h}{\epsilon_h},
\end{eqnarray}
where the ultraviolet cutoff $\epsilon_h$ and the infrared cutoff
$\Lambda_h$ have been set by $\eta_h^2=2\epsilon_h^2/15$ and
$\tilde{N}=\Lambda_h^2/\epsilon_h^2$ \cite{rbm,jlj}, the proper
distance $\eta_h$ from the event horizon to the inner brick wall is
$\eta_H=\int_{U_H}^{U_H+\tilde{\varepsilon}}
\sqrt{-g_{UU}+g^2_{VU}/g_{VV}}dU \thickapprox 2\sqrt{\varepsilon
r_H/(1-\lambda r_H^2)}$ and from the cosmological horizon to the
brick wall is $\eta_C \thickapprox 2 \sqrt{\varepsilon
r_C/(1-\lambda r_C^2)}$, and $A_h=4\pi r_H^2$ or $4\pi r_C^2$.

Comparing with Eqs. (\ref{entropy}) and (\ref{entropy2}), we find
that it equals to  the  entropies calculated in the Painlev\'e and
Schwarzschild-like coordinates.

From above discussions we find that although both the Painlev\'e and
the Lemaitre spacetimes do not possess the singularity at the event
and cosmological horizons, the entropies calculated in the
Painlev\'e and the Lemaitre coordinates are equivalent to that
calculated in the Schwarzschild-like coordinate. It is well known
that the wave modes obtained by using semiclassical techniques, in
general, are the exact modes of the quantum system in the asymptotic
regions. Thus, if the asymptotic structure of the spacetime is the
same for the two coordinates, then the semiclassical wave modes
associated with these two coordinate systems will be the same. From
Eq. (\ref{lem}) we know that the differential relationship between
the Lemaitre time $V$ and the Painlev\'e time $t$ can be expressed
as $dV=dt+d\tilde{r}=2dt+dr/\sqrt{1-f(r)}$. Now let us also work
along the curve $dr+\sqrt{1-f(r)}dt=0$, we obtain $dV=dt$. It is
shown that the two definitions of positive frequency---with respect
to $V$ in the Lemaitre spacetime and with respect to $t$ en the
Painlev\'e spacetime---do coincide. Therefore, it should not be
surprised at the  entropies driven from the modes in the Lemaitre
and Painlev\'e coordinates are the same.
 \vspace*{0.4cm}

\section{Summary}

We have studied the statistical-mechanical entropies  arising from
the quantum massless arbitrary spin fields in the Painlev\'e and
Lemaitre coordinate representations of the Schwarzschild-de Sitter
black hole using the improved thin-layer BWM. At first sight, we
might have anticipated that the results are different from that of
the Schwarzschild-like coordinate due to two reasons: (a) both the
Painlev\'e and Lemaitre coordinate representations possess a
distinguishing property---there are no singularities at $f(r)=0$ so
the metrics are regular at the event and cosmological horizons of
black hole; (b) it is not obvious that the time $V$ in the Lemaitre
spacetime tends to the time $t$ in the Painlev\'e spacetime.
However, by comparing our results (\ref{entropy2}) and
(\ref{entropy3}), which are worked out exactly, with the well-known
result (\ref{entropy}), we have found that in both these coordinate
representations the entropies are the same as that in the standard
Schwarzschild-like coordinate representation.

There are two reasons lead us to obtain the same results in the
different coordinates. a) Although either the Painlev\'e or Lemaitre
coordinate does not possess the singularity, the event  and
cosmological horizons manifests themselves as singularities in the
action function and then there could be particles production. Hence
we can use the knowledge of the wave modes of the quantum field to
calculate the statistical-mechanical entropies. b) When we construct
a vacuum state for the massless arbitrary spin fields in the
Painlev\'e spacetime we take the condition $dr+\sqrt{1-f(r)}dt=0$,
and then we find that the modes used to calculate the entropies in
both the Painlev\'e and Lemaitre coordinates are exactly the same as
that in the Schwarzschild-like coordinates since both $V$ and $t$
tend to the Schwarzschild time $t_s$ as $r$ goes to infinity under
this condition. Therefore, it should not be a surprise that the
entropies driven from the modes in the Lemaitre, Painlev\'e, and
Schwarzschild-like coordinates are the same.

\begin{acknowledgments}This work was supported by the
National Natural Science Foundation of China under Grant No.
10675045; the FANEDD under Grant No. 200317; and the SRFDP under
Grant No. 20040542003.
\end{acknowledgments}


\begin{thebibliography}{99}


\bibitem{be} J. D. Bekenstein, Lett. Nuovo Cimento Soc.
Ital. Fis. {\bf 4}, 737 (1972);
  {\it Black Holes and Entropy} Phys. Rev. D {\bf 7}, 2333 (1973).
\bibitem{ha} S. W. Hawking, {\it Black hole explosions?}  Nature (London) {\bf 248}, 30 (1974);
{\it Particle creation by black holes} Commun. Math. Phys. {\bf 43},
199 (1975).

\bibitem{Carlip} S. Carlip, {\it Black Hole Entropy from Conformal Field Theory in Any Dimension}
 Phys. Rev. Lett. 82, 2828 (1999).

\bibitem{CAr} S. Carlip, {\it Entropy from Conformal field theory at Killing horizens}
 Class. Quantum Grav. 16, 3327 (1999).

\bibitem{JLJing} Jiliang Jing and Mulin Yan, \textit{Statistical
entropy of a stationary dilaton black hole from the Cardy formula}
Phys.  Rev. D, {\bf 63}, (2002) 024003.


\bibitem{sqw} S. Q. Wu and M. L. Yan, {\it Entropy of a Kerr-de Sitter black hole due to arbitrary spin fields}
 Phys. Rev. D {\bf 69},
044019 (2004).
\bibitem{jing} J. L. Jing, {\it Can the ``brick wall" model present
 the same results in different coordinate representations?}
 Phys. Rev. D {\bf69}, 024011 (2004).
\bibitem{chen} J. L. Jing and S. B. Chen, {\it Entropy of the Schwarzschild Black Hole in the Painleve and the
Lemaitre Coordinates} Chin. Phys. Lett. {\bf 21}, 432 (2004).
\bibitem{agh} A. Ghosh and
P. Mitra, {\it Entropy in Dilatonic Black Hole Background }  Phys.
Rev. Lett. {\bf73}, 2521 (1994); {\it Entropy for extremal
Reissoner-Nordstrom black holes} Phys. Lett. B {\bf357}, 295 (1995).
 \bibitem{hsu}  H. Suzuki,
 E. Takasugi and H. Umetsu, {\it Perturbations of Kerr-de Sitter
 Black Holes and Heun's Equations} Prog. Theor. Phys. {\bf100}, 491 (1998);
  {\it Analytic Solutions of the Teukolsky Equation in Kerr-de Sitter and
 Kerr-Newman-de Sitter Geometries}
 {\bf102}, 253 (1999).

 \bibitem{mhl}  M. H. Lee and J. K. Kim, {\it Entropy of a quantum field in rotating black holes } Phys. Rev. D
{\bf 54}, 3904 (1996); {\it The entropy of a quantum field in a
charged Kerr black hole} Phys. Lett. A {\bf 212}, 323 (1996).
\bibitem{rbm} R. B. Mann and S. N. Solodukhin, {\it conical geometry and quantum entropy of a charged kerr
black hole} Phys. Rev. D {\bf54}, 3932 (1996).
\bibitem{jlj}
J. L. Jing and M. L. Yan, {\it Effect of spin on the quantum entropy
of black holes} Phys. Rev. D {\bf63}, 084028 (2001); {\it Quantum
entropy of Kerr the black hole arising from gravitational
perturbation }
 {\bf64}, 064015 (2001); J. L. Jing, {\it Quantum Coorection to Entropy
 of the Kerr Black Hole due Rarita-Schwinger Fields} Chin. Phys. Lett. {\bf20},
 459 (2003).

\bibitem{Jil}  J. L. Jing and  M. L. Yan, {\it Quantum entropy of a nonextreme
ststionary axisymmetric black hole due to a minimally coupled
quantum scalar fields} Phys. Rev. D {\bf 60}, 084015 (1999); {\it
Entropies of the general nonextreme stationary axisymmetric black
holes: Statistical mechanics and thermodynamics} {\bf 61}, 044016
(2000).
\bibitem{sqw3}

S. Q. Wu and X. Cai, {\it Hawking Radiation of Dirac Particles in a
Variable-Mass Kerr Black Hole} Chin. Phys. Lett. {\bf18}, 485
(2001); {\it Hawking Radiation of Dirac Particles in a Variable-Mass
Kerr Space-time } Gen. Relativ. Gravit. {\bf33}, 1181 (2001); {\it
Addendum: Hawking Radiation of a Photons in a Variable-Mass Kerr
Black Hole} {\it ibid}. {\bf34}, 557 (2002), {\it Hawking Radiation
of a nonstationary Kerr-Newman Black Hole: Spin-Rotation Coupling
Effect} {\bf34},605 (2002).
\bibitem{cjg} C. J. Gao and W. B. Liu, {\it Entropy of the Dirac field
in the Schwarchild-de Sitter Spacetime Via Membrane Model} Int. J.
Theor. Phys. {\bf39}, 2221 (2000);
 W. B. Liu and Z. Zhao, Int. J. Mod. Phys. A {\bf16}, 3793 (2001);
{\it An Improved Thin Film Brick Wall Model of Black Hole Entropy}
Chin. Phys. Lett. {\bf18}, 310 (2001).
\bibitem{xli} X. Li and Z. Zhao, {\it Entropy of a Vaidya black hole}
 Phys. Rev. D {\bf62}, 104001 (2000); F. He, Z. Zhao, and S. W.
 Kim, {\it Statistical entropies of scalar and spinor fields in Vaidya-de Sitter
 space-time computed by the thin-layer methord }
{\it ibid}. {\bf64}, 044025  (2001).

\bibitem{G.} G 't. Hooft, {\it On the quantum structure of a black hole}  Nucl. Phys. B {\bf 256}, 727 (1985).

\bibitem{sh1} S. Shankaranarayanan and K. Srinivasan and T. Padmanabhan, {\it METHORD OF COMPLEX PATH AND
GENERAL CONVARIANCE OF HAWKING RADIATION} Mod. Phys. Lett. A
{\bf16}, 571 (2001).
\bibitem{sh2} S. Shankaranarayanan and T. Padmanabhan and K. Srinivasan,
{\it Hawking radiation in different coordinate settings: complex
paths approach} Class. Quantum Grav. {\bf19}, 2671 (2002).

\bibitem{en}  E. Newman and R. Penrose, {\it An approach to gravitational
radiation by a method of spin coefficients} J. Math. Phys. {\bf3},
566 (1962).
\bibitem{sch}
 S. Chandrasekhar, {\it The Mathematical Theory of Black
Holes} (Oxford University Press, New York, 1983).

\bibitem{sat}
A. TeukolskyS, {\it Rotating Black Holes: Separable Wave Equations
for Gravitational and Electromagnetic Perturbations } Phys. Rev.
Lett. {\bf29}, 1114 (1972); Astrophys. J. {\bf185}, 635 (1973); W.
H. Press and S. A. Teukolsky {\it ibid}. {\bf185}, 649 (1973); S. A.
Teukolsky and W. H. Press {\it ibid}. {\bf193}, 443 (1974).
 \bibitem{gft}  G. F. Torres del Castillo, {\it The Teukolsky-Starobinsky
 identities in type D vacuum backgrounds with cosmological constant }  J.
Math. Phys. {\bf29}, 2078 (1988); {\it Rarita-Schwinger fields in
argebraically apecial vacuum space-time} {\bf30}, 446 (1989); {\it
Debye potatials for Rarita-Schwinger fields in curved space-time}
{\bf30}, 1323 (1989).


\bibitem{pkra} P. Kraus, and Frank Wilczek, {\it Some Applications of a Simple Stationary Line Element
for the Schwarzschild Geometry} gr-qc/9406042.

\end{thebibliography}
\end{document}